\documentclass{LMCS}

\usepackage{url}

\usepackage{amssymb} 
\usepackage{amsmath}  
\usepackage{amsbsy} 
\usepackage{amsfonts} 
\usepackage{euscript} 
\usepackage{graphicx} 
\usepackage{stmaryrd} 
\usepackage{enumerate,hyperref}

\theoremstyle{plain}
\theoremstyle{plain}
\theoremstyle{plain}\newtheorem{proposition}[thm]{Proposition}
\theoremstyle{plain}

\theoremstyle{definition}

\def\eqalign#1{\null\,\vcenter{\openup\jot\mathsurround=0 pt
  \ialign{\strut\hfil$\displaystyle{##}$&$\displaystyle{{}##}$\hfil
      \crcr#1\crcr}}\,}

\def\doi{5 (1:7) 2009}
\lmcsheading%
{\doi}
{1--19}
{}
{}
{Mar.~13, 2008}
{Mar.~31, 2009}
{}   

\begin{document}

\title{On tiered small jump operators}

\author[J.-Y.~Marion]{Jean-Yves Marion}
\address{Nancy Universit\'e, Loria, INPL-ENSMN, B.P. 239, 
54506 Vand\oe  uvre-l\`es-Nancy Cedex, France, 
France.}
\email{Jean-Yves.Marion@loria.fr}  
%\thanks{thanks 1, optional.}   %optional

%\titlecomment{OPTIONAL comment concerning the title, \eg, if a variant
%or an extended abstract of the paper has appeared elsewehere}
\maketitle

%%%%
% Macros
%%%

% type 
\newcommand{\pW}{\mathcal{W}}

%%% Sémantique %%%
\newcommand{\sem}[1]{\llbracket #1 \rrbracket}
%%%
\newcommand{\N}{\mathbb{N}}

% domaine
\newcommand{\sW}{\mathbb{W}}
% fonctions
\newcommand{\suca}[1]{A_{#1}}
\newcommand{\sucb}[1]{B_{#1}}
\newcommand{\suc}[1]{S_{#1}}
\newcommand{\mv}[1]{0_{#1}}

\newcommand{\pair}[3]{\langle #2,#3 \rangle_{#1}}
\newcommand{\pra}[1]{\pi^{1}_{#1}}
\newcommand{\prb}[1]{\pi^{2}_{#1}}

\newcommand{\RI}[2]{\mathcal{L}_{#2}(#1)}
\newcommand{\SRI}[2]{\mathcal{I}_{#2}(#1)}

\newcommand{\FP}{\text{FPTIME}}
\newcommand{\DT}{\text{DTIME}}

\newcommand{\sra}{\rightarrow}

\newcommand{\fadd}[1]{\textit{add}_{#1}}
\newcommand{\fmult}[1]{\textit{mul}_{#1}}
\newcommand{\fsm}[1]{\textit{m}_{#1}}
\newcommand{\cinq}{\textit{five}}
\newcommand{\fcube}[1]{\textit{cube}_{#1}}
\newcommand{\ftrois}[1]{\textit{cube'}_{#1}}
\newcommand{\fdeux}[1]{\textit{mul'}_{#1}}

\newcommand{\lab}{\textsc{label}}
\newcommand{\pred}{\textsc{pred}}
\newcommand{\branch}{\textsc{branch}}
\newcommand{\Reg}{\mathcal{R}}
\newcommand{\State}{\mathcal{S}}
\newcommand{\store}{\sigma}

\newcommand{\efface}[1]{\kappa_{#1}}
\newcommand{\coerce}[1]{\textit{coerce}_{#1}}

\newcommand{\itd}[1]{\Delta[#1]}
\newcommand{\itdk}[2]{\Delta^{#1}[#2]}
\newcommand{\itdiag}{\Delta^{\omega}}
\newcommand{\bkt}[1]{\mathcal{B}[{#1}]}

% typed system
\newcommand{\dom}{\text{dom}}
\newcommand{\LD}{\boldsymbol{\lambda}\mathbf{D}^{\boldsymbol{\omega}}}
\newcommand{\bkd}{*}
\newcommand{\kdi}{\Rightarrow}
\newcommand{\kindterm}{\kappa}
\newcommand{\atier}{\omega}
\newcommand{\tW}{\mathbf{W}}
\newcommand{\tzero}{\mathbf{\diamond}} % ! 0
\newcommand{\tsuc}{\textbf{S}}
\newcommand{\tmv}{\boldsymbol{0}} % ! \epsilon
\newcommand{\tsuca}{\mathbf{A}}
\newcommand{\tsucb}{\mathbf{B}}
\newcommand{\tsucj}{\mathbf{J}}
\newcommand{\tsuci}{\mathbf{I}}
\newcommand{\tpair}[3]{\boldsymbol{\langle} #1,#2 \boldsymbol{\rangle}_#3}
\newcommand{\tpra}{\boldsymbol{\pi}^1}
\newcommand{\tprb}{\boldsymbol{\pi}^2}
\newcommand{\tflat}{\textbf{flat}}
\newcommand{\diag}{\textbf{diag}}

\newcommand{\imp}{\vdash}
\newcommand{\typenv}{\Gamma}
\newcommand{\typone}{\tau}
\newcommand{\termone}{M}
\newcommand{\termtwo}{N}
\newcommand{\conone}{\mathbf{c}}
\newcommand{\varone}{x}

\newcommand{\tr}[1]{\underline{#1}}
\newcommand{\trw}[2]{\underline{#2}_{#1}}
\newcommand{\CI}[2]{\mathcal{CI}_{#1}(#2)}

\newcommand{\red}{\rhd}
\newcommand{\redtr}{\red^*}

%%%%%%%%%%%%%%%%%%%%%%%%%%%%%%%%%%%%%%%%%%%%%%%%%%%%%%%%%%%%%%%%%%%%%%
%
%     Prooftree
%
%%%%%%%%%%%%%%%%%%%%%%%%%%%%%%%%%%%%%%%%%%%%%%%%%%%%%%%%%%%%%%%%%%%%%%

\message{<Paul Taylor's Proof Trees, 2 August 1996>}

\def\introrule{{\cal I}}\def\elimrule{{\cal E}}%%
\def\andintro{\using{\land}\introrule\justifies}%%
\def\impelim{\using{\Rightarrow}\elimrule\justifies}%%
\def\allintro{\using{\forall}\introrule\justifies}%%
\def\allelim{\using{\forall}\elimrule\justifies}%%
\def\falseelim{\using{\bot}\elimrule\justifies}%%
\def\existsintro{\using{\exists}\introrule\justifies}%%

%% #1 is meant to be 1 or 2 for the first or second formula
\def\andelim#1{\using{\land}#1\elimrule\justifies}%%
\def\orintro#1{\using{\lor}#1\introrule\justifies}%%

%% #1 is meant to be a label corresponding to the discharged hypothesis/es
\def\impintro#1{\using{\Rightarrow}\introrule_{#1}\justifies}%%
\def\orelim#1{\using{\lor}\elimrule_{#1}\justifies}%%
\def\existselim#1{\using{\exists}\elimrule_{#1}\justifies}

%%==========================================================================

\newdimen\proofrulebreadth \proofrulebreadth=.05em
\newdimen\proofdotseparation \proofdotseparation=1.25ex
\newdimen\proofrulebaseline \proofrulebaseline=2ex
\newcount\proofdotnumber \proofdotnumber=3
\let\then\relax
\def\hfi{\hskip0pt plus.0001fil}
\mathchardef\squigto="3A3B
%
% flag where we are
\newif\ifinsideprooftree\insideprooftreefalse
\newif\ifonleftofproofrule\onleftofproofrulefalse
\newif\ifproofdots\proofdotsfalse
\newif\ifdoubleproof\doubleprooffalse
\let\wereinproofbit\relax
%
% dimensions and boxes of bits
\newdimen\shortenproofleft
\newdimen\shortenproofright
\newdimen\proofbelowshift
\newbox\proofabove
\newbox\proofbelow
\newbox\proofrulename
%
% miscellaneous commands for setting values
\def\shiftproofbelow{\let\next\relax\afterassignment\setshiftproofbelow\dimen0 }
\def\shiftproofbelowneg{\def\next{\multiply\dimen0 by-1 }%
\afterassignment\setshiftproofbelow\dimen0 }
\def\setshiftproofbelow{\next\proofbelowshift=\dimen0 }
\def\setproofrulebreadth{\proofrulebreadth}

%=============================================================================
\def\prooftree{% NESTED ZERO (\ifonleftofproofrule)
%
% first find out whether we're at the left-hand end of a proof rule
\ifnum  \lastpenalty=1
\then   \unpenalty
\else   \onleftofproofrulefalse
\fi
%
% some space on left (except if we're on left, and no infinity for outermost)
\ifonleftofproofrule
\else   \ifinsideprooftree
        \then   \hskip.5em plus1fil
        \fi
\fi
%
% begin our proof tree environment
\bgroup% NESTED ONE (\proofbelow, \proofrulename, \proofabove,
%               \shortenproofleft, \shortenproofright, \proofrulebreadth)
\setbox\proofbelow=\hbox{}\setbox\proofrulename=\hbox{}%
\let\justifies\proofover\let\leadsto\proofoverdots\let\Justifies\proofoverdbl
\let\using\proofusing\let\[\prooftree
\ifinsideprooftree\let\]\endprooftree\fi
\proofdotsfalse\doubleprooffalse
\let\thickness\setproofrulebreadth
\let\shiftright\shiftproofbelow \let\shift\shiftproofbelow
\let\shiftleft\shiftproofbelowneg
\let\ifwasinsideprooftree\ifinsideprooftree
\insideprooftreetrue
%
% now begin to set the top of the rule (definitions local to it)
\setbox\proofabove=\hbox\bgroup$\displaystyle % NESTED TWO
\let\wereinproofbit\prooftree
%
% these local variables will be copied out:
\shortenproofleft=0pt \shortenproofright=0pt \proofbelowshift=0pt
%
% flags to enable inner proof tree to detect if on left:
\onleftofproofruletrue\penalty1
}

%=============================================================================
% end whatever box and copy crucial values out of it
\def\eproofbit{% NESTED TWO
%
% various hacks applicable to hypothesis list 
\ifx    \wereinproofbit\prooftree
\then   \ifcase \lastpenalty
        \then   \shortenproofright=0pt  % 0: some other object, no indentation
        \or     \unpenalty\hfil         % 1: empty hypotheses, just glue
        \or     \unpenalty\unskip       % 2: just had a tree, remove glue
        \else   \shortenproofright=0pt  % eh?
        \fi
\fi
%
% pass out crucial values from scope
\global\dimen0=\shortenproofleft
\global\dimen1=\shortenproofright
\global\dimen2=\proofrulebreadth
\global\dimen3=\proofbelowshift
\global\dimen4=\proofdotseparation
\global\count255=\proofdotnumber
%
% end the box
$\egroup  % NESTED ONE
%
% restore the values
\shortenproofleft=\dimen0
\shortenproofright=\dimen1
\proofrulebreadth=\dimen2
\proofbelowshift=\dimen3
\proofdotseparation=\dimen4
\proofdotnumber=\count255
}

%=============================================================================
\def\proofover{% NESTED TWO
\eproofbit % NESTED ONE
\setbox\proofbelow=\hbox\bgroup % NESTED TWO
\let\wereinproofbit\proofover
$\displaystyle
}%
%
%=============================================================================
\def\proofoverdbl{% NESTED TWO
\eproofbit % NESTED ONE
\doubleprooftrue
\setbox\proofbelow=\hbox\bgroup % NESTED TWO
\let\wereinproofbit\proofoverdbl
$\displaystyle
}%
%
%=============================================================================
\def\proofoverdots{% NESTED TWO
\eproofbit % NESTED ONE
\proofdotstrue
\setbox\proofbelow=\hbox\bgroup % NESTED TWO
\let\wereinproofbit\proofoverdots
$\displaystyle
}%
%
%=============================================================================
\def\proofusing{% NESTED TWO
\eproofbit % NESTED ONE
\setbox\proofrulename=\hbox\bgroup % NESTED TWO
\let\wereinproofbit\proofusing
\kern0.3em$
}

%=============================================================================
\def\endprooftree{% NESTED TWO
\eproofbit % NESTED ONE
% \dimen0 =     length of proof rule
% \dimen1 =     indentation of conclusion wrt rule
% \dimen2 =     new \shortenproofleft, ie indentation of conclusion
% \dimen3 =     new \shortenproofright, ie
%                space on right of conclusion to end of tree
% \dimen4 =     space on right of conclusion below rule
  \dimen5 =0pt% spread of hypotheses
% \dimen6, \dimen7 = height & depth of rule
%
% length of rule needed by proof above
\dimen0=\wd\proofabove \advance\dimen0-\shortenproofleft
\advance\dimen0-\shortenproofright
%
% amount of spare space below
\dimen1=.5\dimen0 \advance\dimen1-.5\wd\proofbelow
\dimen4=\dimen1
\advance\dimen1\proofbelowshift \advance\dimen4-\proofbelowshift
%
% conclusion sticks out to left of immediate hypotheses
\ifdim  \dimen1<0pt
\then   \advance\shortenproofleft\dimen1
        \advance\dimen0-\dimen1
        \dimen1=0pt
%       now it sticks out to left of tree!
        \ifdim  \shortenproofleft<0pt
        \then   \setbox\proofabove=\hbox{%
                        \kern-\shortenproofleft\unhbox\proofabove}%
                \shortenproofleft=0pt
        \fi
\fi
%
% and to the right
\ifdim  \dimen4<0pt
\then   \advance\shortenproofright\dimen4
        \advance\dimen0-\dimen4
        \dimen4=0pt
\fi
%
% make sure enough space for label
\ifdim  \shortenproofright<\wd\proofrulename
\then   \shortenproofright=\wd\proofrulename
\fi
%
% calculate new indentations
\dimen2=\shortenproofleft \advance\dimen2 by\dimen1
\dimen3=\shortenproofright\advance\dimen3 by\dimen4
%
% make the rule or dots, with name attached
\ifproofdots
\then
        \dimen6=\shortenproofleft \advance\dimen6 .5\dimen0
        \setbox1=\vbox to\proofdotseparation{\vss\hbox{$\cdot$}\vss}%
        \setbox0=\hbox{%
                \advance\dimen6-.5\wd1
                \kern\dimen6
                $\vcenter to\proofdotnumber\proofdotseparation
                        {\leaders\box1\vfill}$%
                \unhbox\proofrulename}%
\else   \dimen6=\fontdimen22\the\textfont2 % height of maths axis
        \dimen7=\dimen6
        \advance\dimen6by.5\proofrulebreadth
        \advance\dimen7by-.5\proofrulebreadth
        \setbox0=\hbox{%
                \kern\shortenproofleft
                \ifdoubleproof
                \then   \hbox to\dimen0{%
                        $\mathsurround0pt\mathord=\mkern-6mu%
                        \cleaders\hbox{$\mkern-2mu=\mkern-2mu$}\hfill
                        \mkern-6mu\mathord=$}%
                \else   \vrule height\dimen6 depth-\dimen7 width\dimen0
                \fi
                \unhbox\proofrulename}%
        \ht0=\dimen6 \dp0=-\dimen7
\fi
%
% set up to centre outermost tree only
\let\doll\relax
\ifwasinsideprooftree
\then   \let\VBOX\vbox
\else   \ifmmode\else$\let\doll=$\fi
        \let\VBOX\vcenter
\fi
% this \vbox or \vcenter is the actual output:
\VBOX   {\baselineskip\proofrulebaseline \lineskip.2ex
        \expandafter\lineskiplimit\ifproofdots0ex\else-0.6ex\fi
        \hbox   spread\dimen5   {\hfi\unhbox\proofabove\hfi}%
        \hbox{\box0}%
        \hbox   {\kern\dimen2 \box\proofbelow}}\doll%
%
% pass new indentations out of scope
\global\dimen2=\dimen2
\global\dimen3=\dimen3
\egroup % NESTED ZERO
\ifonleftofproofrule
\then   \shortenproofleft=\dimen2
\fi
\shortenproofright=\dimen3
%
% some space on right and flag we've just made a tree
\onleftofproofrulefalse
\ifinsideprooftree
\then   \hskip.5em plus 1fil \penalty2
\fi
}

%==========================================================================
% IDEAS
% 1.    Specification of \shiftright and how to spread trees.
% 2.    Spacing command \m which causes 1em+1fil spacing, over-riding
%       exisiting space on sides of trees and not affecting the
%       detection of being on the left or right.
% 3.    Hack using \@currenvir to detect LaTeX environment; have to
%       use \aftergroup to pass \shortenproofleft/right out.
% 4.    (Pie in the sky) detect how much trees can be "tucked in"
% 5.    Discharged hypotheses (diagonal lines).
\newcommand{\infer}[2]
     {\prooftree
          #1 
          \justifies #2
      \endprooftree}

\newcommand{\ninfer}[3]
     {\prooftree
          #1 
          \justifies #2
          \using #3
      \endprooftree}

%\hyphenation{qua-si intre-pre-tation inter-pre-tations}

\newcommand{\modifJY}[1]{\textcolor{blue}{#1}}
\newcommand{\modifR}[1]{\textcolor{red}{#1}}

\begin{abstract} 
Predicative analysis of recursion schema is a method to characterize
complexity classes like the class $\FP$ of polynomial time computable
functions. This analysis comes from the works of Bellantoni and Cook,
and Leivant by data tiering.  Here, we refine predicative analysis by
using a ramified Ackermann's construction of a non-primitive recursive
function. We obtain a hierarchy of functions which characterizes
exactly functions, which are computed in $O(n^k)$ time over register
machine model of computation.  For this, we introduce a strict
ramification principle.  Then, we show how to diagonalize in order to
obtain an exponential function and to jump outside deterministic
polynomial time.
%$\cup_k \DT(n^k)$.
Lastly, we suggest a dependent typed lambda-calculus to represent this
construction.
\end{abstract}

\section{Introduction}
Predicative analysis of recursion comes from the works of Bellantoni
and Cook \cite{BC92} and Leivant \cite{Lei94}. This analysis is based
on a ramification principle on data which is appealing because its
concept is simple and purely syntactic and does not involve parts of
its models.  Each element of a computation has a tier, which
determines its ability to run a recursion. The ramification principle
states that a definition by recursion is ramified only if the tier of
the recurrence parameter is strictly higher than the tier of the
output. This analysis takes its root in the paper of Simmons
\cite{Sim88} and Leivant \cite{Lei91}.  The results mentioned above
characterize the class of polynomial time computable functions using
essentially two tiers of data ramification: one for recursion
arguments and one for recursion outputs.  In this work, we revisit the
ramification principle by introducing a \emph{strict ramification
  principle} which allows getting a characterization of a polynomial
time hierarchy of functions. Functions which are defined with $k$
tiers are exactly functions which are computable in $O(n^k)$ steps.
The hierarchy is not robust in the sense that it depends on the model
of computation which is a register machine model here. So, the result
that we suggest is really about intrinsic complexity of functions in
the tradition of the recursion Theory.  We have tried to understand
the mechanism that underpins the suggested classification. Our
analysis shows how functions are defined and how we can jump from one
class of functions to another one by strict ramified recursion. This
leads us to introduce a double recursion operator, which captures each
level of the polynomial time hierarchy $\DT(n^k)$ and escapes
them. For this, we define an exponential function by a diagonalization
method, which reveals some analogies with Ackermann \cite{Ack28}
construction as it is explained in Chapter $7$ of Simmons book
\cite{Sim00}. The construction that we propose is a kind of double
recursion whose main ideas can be explained by considering the
following example.\smallskip

\[f:\N(1),\N(0) \sra \N(0)\qquad
\eqalign{
f(0,y)  & = y+1 \cr
f(x+1,y) & = f(x,f(x,y))
  }
\]\smallskip

\noindent The function $f$ is defined by nested recursion and
satisfies the ramification principle. Indeed, the first argument may
be of tier $1$ and the second of tier $0$. So, the output of $f$ is of
tier $0$ and $f$ is well typed. However $f$ computes the exponential
function : $f(n,m)=2^n + m$ for all $n$ and $m$. In $f(x,f(x,y))$, the
leftmost occurrence of $f$ calls itself which violates the essence of
the ramification principle. Now, we ramify $f$ by assigning to each
occurrence of $f$ a tier, and so we obtain the following function
sequence.
\begin{align*}
f_0(x,y)  & = y+1 \\
f_{k+1}(0,y) & = y \\
f_{k+1}(x+1,y) & = f_{k}(x,f_{k+1}(x,y)) 
\end{align*} 
where $f_1$ computes the addition, and $f_2$ iterates the addition,
and so on.  We also see that the domain, or the type, of each $f_k$ can be
$\N(k),\N(0) \sra \N(0)$. If we transform the $(f_k)_{k\in\N}$
sequence of functions into a three place function $\phi(k,x,y)$, we
are able to produce by a diagonalization argument a function which
eventually dominates each $f_k$. The type of $\phi$ depends on its
first argument and so would be $\forall k:\N(k),\N(0) \sra \N(0)$. 

This example is just here to illustrate quickly the ideas that we
develop in this paper, which is organized as follows. 
Section 2 presents the computational models and defines
$\DT(n^k)$. Section 3 focuses on tiered recursion and Leivant's
characterization of $\FP$. This Section contains well-known material, and so the paper is self-contained.
 Section 4 gives the characterization of the
polynomial time hierarchy. Section 5 describes how to jump from
$\DT(n^k)$ to $\DT(n^{k+1})$ and how to diagonalize in order to escape
$\FP$. In the last section, an applied typed lambda-calculus, like in
Simmons survey~\cite{Simmons05}, with dependent types is proposed to
represent the jump operator presented in the previous Section.

%We choose except for the last part to present this work in term of
%tiered function algebra. It is not too difficult to translate this formalism into an
%applied typed lambda-calculus, like in Simmons survey~\cite{Simmons05}. 
%(and we partially do it in Section 5.)
%Lastly, we make this choice because it is more readable and so easier to
%understand.

\section{Computations and a polynomial time hierarchy}
\subsection{Register Machines}
The set of binary words over the alphabet $\{a,b\}$ is $\sW$.
A register machine, abbreviated RM, works over words of $\sW$.
A RM consists in
\begin{enumerate}[(1)]
\item an alphabet $\{a,b\}$. 
\item a finite set $\State = \{ s_0, s_1, \ldots, s_k \}$ of states,
  including a distinct state \textsc{begin}.
\item a finite list $\Reg = \{R_1, \ldots, R_m\}$ of
  registers. Registers store words of $\sW$.
\item a finite function $\lab$ mapping states to commands which are 
  \begin{align*}
  R & = a(R) && \textit{add the letter $a$ to $R$} \\
  R & = b(R) && \textit{add the letter $b$ to $R$} \\
  R & = R' && \textit{assign the value of $R'$ to $R$} \\
  R & = \pred(R) && \textit{remove the first letter of $R$} \\
  & \branch(R,s_{\epsilon},s_{a},s_{b})  
     && \textit{switch to the label $s_{i}$ following the first letter of $R$}
  \end{align*}
\end{enumerate}

A \emph{configuration} of a RM $M$ is given by a pair $(s,\store)$ where
$s$ is a state and $\store: \Reg \to \sW$ is an environment which
stores register values.
We guess that the above informal semantics should be enough to understand
how register machines work.
In particular, after executing one of the four first kinds of instruction, if the state is $s_i$ and $i<k$, then the next state is $s_{i+1}$.
Otherwise, if the state is $s_k$, then $M$ halts. Lastly, the next step of a branching instruction depends on the value of the register $R$.

Throughout, we deal with functions which have a co-arity, that is
function whose range is $\sW^{q}$ for some $q$.
A function $\phi:\sW^{p} \sra \sW^{q}$ is computed by a register
machine $M$ if for all $u_1,\ldots,u_{p}$, $p \leq m$
we have $\phi(u_1,\ldots,u_{p}) = (v_1,\ldots,v_{q})$ 
then the execution of $M$ starting from the initial configuration
$(\textsc{begin},\sigma_0)$ ends to a configuration $(s,\sigma_f)$
such that:
for $i \leq p$, $\sigma_0(R_i) = u_i$, otherwise $\sigma_0(R_i)=\epsilon$
and for $j \leq q$, $\sigma_f(R_{m+1-j}) = v_j$.

\subsection{A polynomial time hierarchy}

The time measure corresponds to the number of steps to perform a
computation on a register machine. 
We say that a function $\phi:\sW^{p} \sra \sW^{q}$ is
computable in
$O(n^k)$ if the runtime is bounded by
$c.(n_1^k+\ldots+n_p^k)+d$ for some $c$ and $d$
and where for each $i$, $n_i$ is the size of the $i$\textit{th} argument.
The class $\DT(n^k)$ is the set of all functions which are computable
in $O(n^k)$.
The class $\FP$ of polynomial time functions  is $\cup_k \DT(n^k)$.

In this work, we study the classes $\DT(n^k)$ which delineates a
polynomial time hierarchy. It is well known that the class $\FP$ is robust, which is not the case
for polynomial time hierarchies. Indeed, the definition of $\DT(n^k)$ is
not invariant with respect to another class of computational
models. The reason lies on the fact that the simulation of a computational
model by another may have a quadratic cost. 
For example, the runtime of simulations
of a two-tape Turing machine by a one-tape Turing machine is
quadratic. Such lower bound may be nicely obtained using Kolmogorov complexity. 
The reader may consult Jones' book~\cite{JonesCC} for further informations. 
However, one may use k-tape Turing machines instead of register machines.

\section{Ramified Primitive recursion}
\subsection{Functions on tiered domains}
We are interested in computational complexity, that is why we focus
immediately on words. 
The domain of reference is the set $\sW$ of words over the alphabet
$\{a,b\}$. 
It is generated from the empty word function $\mv{}$  
and two successors $\suca{}$ and
$\sucb{}$. As usual
$\suca{}(\sucb{}(\mv{}))$ is the word $ab$.

This domain is tiered  by duplicating $\sW$ into
$\sW(0), \sW(1),\ldots,\sW(i),\ldots$ 
where each $\sW(i)$ is an identical copy of $\sW$ at tier $i$.
Each domain $\sW(i)$ is a set of words over the alphabet $\{a_i,b_i\}$.
As previously, there are an empty word function $\mv{i}$ 
and two successors $\suca{i}$ and $\sucb{i}$. 
In practice, we define functions by specifying their values with
respect to tiered domain generators.

There are erasing bijections $\efface{k}:\sW(k) \sra \sW$ for each $k$ which just
erase the tier of words. For example, we may represent a function $\phi:\sW \sra \sW$
by
$f:\sW(k) \sra \sW(0)$ for some tier $k$ if for each $u \in \sW(k)$,
$\efface{0}(f(u)) = \phi(\efface{k}(u))$.
In this case, we shall just write $f(u)=\phi(u)$.

%We also consider \emph{downcasting bijections} $\coerce{k+1}:\sW(k+1)\sra
%\sW(k)$ for each $k$ such that 
%$\coerce{k+1}(\mv{k+1}) = \mv{k}$,
%$\coerce{k+1}(\suca{k+1}(x))  = \suca{k}(\coerce{k+1}(x))$
%and
%$\coerce{k+1}(\sucb{k+1}(x))  = \sucb{k}(\coerce{k+1}(x))$.
%Similarly,

We always reason with respect to an implicit downcasting principle, 
which yields that if $u \in \sW(k+1)$ then $u \in \sW(k)$.
Hence, we shall write that $f:\sW(k+1) \sra \sW(0)$ is defined from
$h:\sW(k+1),\sW(k)\sra\sW(0)$ by $f(x)=h(x,x)$ without mentioning that both occurrences of $x$ are not of the same tier. 
\emph{Throughout, we shall reason with respect to erasing bijections and
implicit downcasting without explicitly mentioning them. }

We consider functions with co-arity. 
For this, we construct Cartesian product of domains of same tier. 
We abbreviate $\sW(i) \times \ldots \times \sW(i)$ by
$\sW(i)^{p}$.
We have a pairing function $\pair{i}{~}{~}$ and
both projections $\pra{i}$ and $\prb{i}$, for each tier $i$.

We often leave out some brackets using familiar conventions
and hence we abbreviate  $\tau_1 \sra ( \ldots ( \tau_n \sra \tau))$ by 
$\tau_1, \ldots,\tau_n \sra \tau$.
It is also convenient to have a normal presentation of functions, that we shall always use.
We shall write 
$f:\sW(i_1)^{p_1},\ldots,\sW(i_n)^{p_n} \sra \sW(r)^{q}$
for an $n$-placed function in such a way that $i_1 \geq \ldots \geq i_n \geq r$. 
We say that the tier of the j\textit{th} argument of $f$ is
$i_j$, and the output tier is $r$. 
We write $\vec{y}$ to mean $y_1,\ldots,y_{n}$ where $y_i$ 
is an element of $\sW(i_j)^{p_j}$.
The size $|u|$ is the number of letters of the word $u$. In particular
the size of the empty word $\epsilon$ is $0$. The size of pair of
words is inductively defined as follows: $|\pair{i}{u}{v}|=|u|+|v|$
at any tier $i$.

Conventions that we have described here will be extended to the typed
lambda calculus that we suggest at the end in a natural manner.

\subsection{Ramified primitive recursion} ~\\
A function 
$f:\sW(k+1),\sW(i_1)^{p_1},\ldots,\sW(i_n)^{p_n} \sra \sW(r)^{q}$ 
is obtained by
\emph{ramified primitive recursion}
from the functions \\
$h_{\epsilon}:\sW(i_1)^{p_1},\ldots,\sW(i_n)^{p_n} \sra \sW(r)^{q}$
and \\
$h_{a},h_{b}: \sW(i_1)^{p_1},\ldots,\sW(i_n)^{p_n},\sW(r)^{q} \sra \sW(r)^{q}$
if 
\begin{align}
f(\mv{k+1},\vec{y}) & = h_{\epsilon}(\vec{y}) \\
f(\suca{k+1}(x),\vec{y}) & = h_{a}(\vec{y},f(x,\vec{y})) \\
f(\sucb{k+1}(x),\vec{y}) & = h_{b}(\vec{y},f(x,\vec{y}))
\end{align}
where conditions $k+1 \geq i_j$ for any $j \leq n$ and $k \geq r$ hold.
We call these last conditions the \emph{ramification principle} based
on Leivant's~\cite{Lei94}.
The first argument is named the recursion argument and its tier is $k+1$.
The ramification principle says that the recurrence tier $k+1$ is strictly greater than the output tier $r$. 

\subsection{Ramified arithmetic}~\label{sec:RA}
In order to compare function growth rate and to illustrate key
notions, it is convenient to have an encoding of natural numbers.
This encoding will be used in Sections~\ref{sec:strict} and~\ref{sec:jump}.

We represent natural numbers by considering
both successors $\suca{i}$ and $\sucb{i}$ as the same. 
Hence, we have a single successor that we write $\suc{i}$, for each tier $i$. 
It should be clear that this encoding is non-injective,
which is sufficient because we are just interested
in the size of the handling values. 
So in this representation, a word represents a natural number, which
corresponds to its size.
Hence, $\mv{i}$ will refer to zero at tier $i$, 
and $S_i(x)$ intuitively increases the size of $x$ by one, 
which corresponds exactly to the successor operation in unary notation.

We represent in \emph{ramified arithmetic} 
an arithmetical function $\phi:\N^{p}\sra \N$
by a function $f:\sW(i_1), \ldots, \sW(i_{p}) \sra\sW(r)$ 
if 
\begin{align*}
\phi(n_1,\ldots,n_{p}) & = |f(u_1,\ldots,u_{p})|  
&& 
\text{for each $u_i$ such that $|u_i|=n_i$ and $i=1,p$}
\end{align*}

Now, we can define below the addition $\fadd{k}$ and the
multiplication $\fmult{k}$ at tier $k$.

\medskip
\noindent
$\fadd{k}:
       \sW(k+1),\sW(k) \sra \sW(k)$   and $|\fadd{k}(u,v)|=|u|+|v|$, for
       all $u$ and $v$.
\begin{align*} 
\fadd{k}(\mv{k+1},y)   & = y \\    
\fadd{k}(\suc{k+1}(x),y) & = \suc{k}(\fadd{k}(x,y)) 
&& \text{where $\suc{i}=\suca{i},\sucb{i}$}\\  
\intertext{$\fmult{k}:
        \sW(k+1),\sW(k+1) \sra \sW(k)$ and 
  $|\fmult{k}(u,v)|=|u|.|v|$, for all $u$ and $v$}
\fmult{k}(\mv{k+1},y)   & = \mv{k} \\   
\fmult{k}(\suc{k+1}(x),y) & = \fadd{k}(y,\fmult{k}(x,y)) 
\end{align*} 

Observe that both arguments of $\fmult{k}$ have the same tier $k+1$.
We may define polynomials by composition from tiered addition and
multiplication, as it is illustrated below.

\medskip
\noindent
$\fcube{k}:\sW(k+2) \sra \sW(k)$ 
\begin{align*} \label{cube}
%\fcube{k}(x) & = \fmult{k}(\fcoerce{k+1}(x),\fmult{k+1}(x,x))   
\fcube{k}(x) & = \fmult{k}(x,\fmult{k+1}(x,x))   
\end{align*}
We see that we compute the arithmetical function $x^3$ by
composing two multiplications. 
However, two copies of the multiplication $\fmult{k}$ and
$\fmult{k+1}$
at different tiers  are necessary.
Notice also that the tier of the first argument, on the right
handside, is lower, which is
possible because of the use of a downcasting bijection. Actually, we
may define $\coerce{k}$ by a simple ramified recursion. 
We may then use it instead of the implicit downcasting,
\noindent
$\coerce{k}: \sW(k+1) \sra \sW(k)$ 
\begin{align*} 
\coerce{k}(\mv{k+1}) & = \mv{k} \\
\coerce{k}(\suc{k+1}(x)) & = \suc{k}(\coerce{k}(x))
\end{align*}
On the other hand, the ramified principle allows also to define a cubic function 
$\ftrois{k}:\sW(k+1)^3 \sra \sW(k)$ using only two tiers as follows:
\begin{align*} 
\fdeux{k}(\mv{k+1},z,t) & = t \\
\fdeux{k}(\suc{k+1}(y),z,t) & = \fadd{k}(z,\fdeux{k}(y,z,t))\\
\ftrois{k}(\mv{k+1},y,z) & = \mv{k+1} \\
\ftrois{k}(\suc{k+1}(x),y,z) & = \fdeux{k}(y,z,\ftrois{k}(x,y,z))
\end{align*}

\subsection{Characterization of \FP}
In 1994, Leivant published an elegant characterization~\cite{Lei94} of $\FP$, which provides
a general framework to study complexity classes. We follow here the
main line of his work. So, we begin by introducing a particular kind
of recursion, named flat recursion.

A function
$f:\sW(r),\sW(i_1)^{p_1},\ldots,\sW(i_n)^{p_n} \sra \sW(r)$ 
is obtained by
\emph{flat recursion} 
from the functions \\
$h_{\epsilon}:\sW(i_1)^{p_1},\ldots,\sW(i_n)^{p_n} \sra \sW(r)$
and \\
$h_{a},h_{b}:\sW(r),\sW(i_1)^{p_1},\ldots,\sW(i_n)^{p_n} \sra \sW(r)$
if
\begin{align}
f(\mv,\vec{y}) & = h_{\epsilon}(\vec{y}) \\
f(\suca{r}(x),\vec{y}) & = h_{a}(x,\vec{y}) \\
f(\sucb{r}(x),\vec{y}) & = h_{b}(x,\vec{y})
\end{align}
This kind of recursion should be viewed as a mere action on the pattern
of the recursive argument. 
Hence and unlike the ramified principle, the tier of a recurrence
argument is not strictly higher that the output tier. 
The use of flat recursion is essential to
define a predecessor over $\sW$ and  conditional functions. 

\begin{defi}
A function $f$ is in $\RI{\sW}{\omega}$ if it is obtained by a finite number
of applications of composition, flat recursion and ramified primitive recursion
beginning with basic functions 
$\mv{k}$, $\suca{k}$,  $\sucb{k}$, $\pair{k}{\_}{\_}$, $\pra{k}$ 
and $\prb{k}$ 
for each tier $k$.
\end{defi}

Leivant demonstrated in~\cite{Lei94} the following result:
\begin{thm}
The class of functions $\RI{\sW}{\omega}$ is exactly the class $\FP$ of the
functions which are polynomial time computable.
\end{thm}

In this presentation we
use functions with co-arity, unlike Leivant which introduces simultaneous ramified
recursion.

Actually, Leivant also showed that only two tiers are
sufficient. More generally, 
\begin{cor}
Let $\RI{\sW}{k}$ be the class of functions restricted over
$\sW(0),\ldots,\sW(k)$. 
For each $k$, the class of functions $\RI{\sW}{k+1}$ is exactly the class $\FP$ of the
functions which are polynomial time computable.
\end{cor}

In the same paper, Leivant shows how to capture $\DT(n^k)$ by
counting the degree of nested recursions. 
%So, he hasn't a calculus to
%characterize $\DT(n^k)$.  

\subsection{Other approaches}
The work of Bellantoni and Cook~\cite{BC92} is similar to the
Leivant's one. They characterize $\FP$ by defining a function algebra in which functions
have two kind of arguments: the normal ones which can be used as
recursion parameters and the safe ones which cannot be used as
recursion parameters. 

As we have seen, only two tiers are necessary to characterize
$\FP$.  Actually, this is also the essence of the characterization by simply typed
lambda calculus of~\cite{LM93}. The tier $1$ arguments are represented
by Church-numerals, and the tier $0$ are represented by constant terms
of atomic type on which no recursion can be made.

\section{Strict ramified primitive recursion}
We present the notion \emph{strict ramified primitive recursion} 
which is central in this study.
A function 
$f:\sW(k),\sW(i_1)^{p_1},\ldots,\sW(i_n)^{p_n} \sra \sW(0)^{q}$ 
is obtained by
\emph{k-ramified recursion} 
from the functions \\
$h_{\epsilon}:\sW(i_1)^{p_1},\ldots,\sW(i_n)^{p_n} \sra \sW(0)^{q}$
and \\
$h_{a},h_{b}:\sW(i_1)^{p_1},\ldots,\sW(i_n)^{p_n},\sW(0)^{q} \sra \sW(0)^{q}$
if 
\begin{align}
f(\mv{k},\vec{y}) & = h_{\epsilon}(\vec{y}) \\
f(\suca{k}(x),\vec{y}) & = h_{a}(\vec{y},f(x,\vec{y})) \\
f(\sucb{k}(x),\vec{y}) & = h_{b}(\vec{y},f(x,\vec{y}))
\end{align}
where the inequalities between tiers $k>i_j$ for each $j \leq n$  and $k>0$ hold. 
We call this last condition \emph{the strict ramification principle}.

\begin{defi}
A function $f$ is in $\SRI{\sW}{k}$ if it is obtained by a finite number
of applications of composition, flat recursion and $i$-ramified
recursion,
beginning with basic functions 
$\mv{i}$, $\suca{i}$,  $\sucb{i}$, $\pair{i}{~}{~}$, $\pra{i}$ 
and $\prb{i}$ for each tier $i \leq k$.
\end{defi}

In particular, a function $\SRI{\sW}{0}$ is not defined by recursion. 
The notion of $1$-ramified recursion was underlying in~\cite{LM99a}, and the notion of $k$-ramification is used in order to characterize the NC$^k$ hierarchy in~\cite{BKMO-CSL08}.

The difference between the ramification principle  and the strict ramification principle is the following:
\begin{enumerate}[(1)]
\item The recursion argument is \emph{strictly} greater than the other argument tiers,
\item and the output tier is $0$.

Otherwise, we could define the function $x^5$ with $0,\ldots,4$ tiers,
that is by $4$-ramified recursion and composition as follows:

\noindent
$\fsm{k}: \sW(k+2),\sW(k+1) \sra \sW(k)$
\begin{align*}
\fsm{k}(\mv{k+2},y)   & = \mv{k} \\   
\fsm{k}(\suc{k+2}(x),y) & = \fadd{k}(y,\fsm{k}(x,y)) 
\intertext{$\cinq:\sW(4)\sra\sW(0)$ and $|\cinq(x)|=|x|^5$}
\cinq(x) & = \fsm{0}(\fsm{2}(x,x),\fsm{1}(x,\fsm{2}(x,x)))
\end{align*}

The fact that the output tier of an recursion is $0$ implies that we
cannot defined  $\coerce{k}$ functions. 
That is why we need to reason modulo downcasting bijections.
\end{enumerate}

\subsection{Strict ramified arithmetic}\label{sec:strict}
We use the same encoding of natural numbers that the one we present 
in Section~\ref{sec:RA} on ramified arithmetic.
However, we slightly modify the way that we represent arithmetical
functions to take into account the fact that outputs are of tier $0$.

An arithmetical function $\phi:\N^{p}\sra \N$
is represented in \emph{strict ramified arithmetic}
by a function $f:\sW(i_1), \ldots, \sW(i_{p}) \sra\sW(0)$ 
if 
\begin{align*}
\phi(n_1,\ldots,n_{p}) & =|f(u_1,\ldots,u_{p})| 
&& \text{for each $u_i$ such that $|u_i|=n_i$, $i=1,p$}
\end{align*}

The addition function defined in Section~\ref{sec:RA} is defined by
$1$-ramified recursion, setting $k=0$.
On the other hand, the definition of the multiplication proposed
in~\ref{sec:RA} does not satisfy the strict ramification principle
because both arguments are of the same tier.

Nevertheless, we can define any polynomial.
For this, we present first a sequence  $(F_k)_{k \in \N}$ of $3$-placed
monotonic functions from an initial $1$-placed function $g:\sW(0)^p
\sra \sW(0)^p$. 
Intuitively, the function $g$ is iterated a number of steps bounded by a polynomial of degree $k$.
This sequence will play a \emph{crucial role} all along the paper.

\medskip
\noindent
$F_{0}:\sW(0),\sW(0),\sW(0)^p \sra \sW(0)^p$
\begin{align*}
F_0(t,x,y) & = g(y) \\
\intertext{$F_{k+1}:\sW(k+1),\sW(k),\sW(0)^p \sra \sW(0)^p$}
F_{k+1}(\mv{k+1},x,y) & = y \\
F_{k+1}(\suc{k+1}(t),x,y) & = F_k(x,x,F_{k+1}(t,x,y))
\end{align*}

It is worth noticing that $(F_k)_{k\in\N}$ is parameterized by the
function $g$. Notice that we use implicitly a downcasting to
lower the tier of the second argument on the right hand side of the
last equation.

\begin{lem}\label{prop:Fk}
For any $k$,$u$,$v$ and $w$, we have
\begin{align*}
F_{k+1}(u,v,w) & = g^{m.n^k}(w) && \text{where $m=|u|$ and $n=|v|$}
\end{align*}
\end{lem}

\proof%begin{proof}
The proof is by induction on $k$.

For $k=0$, we have by recurrence on the size of the first argument :
\begin{align*}
 F_1(0_{1},v,w) & = w \\
 F_1(S_1(t),v,w) & = F_0(v,v,F_1(t,v,w)) \\
                 & = g ( g^{|t|}(w)) = g^{|t|+1}(w)
\end{align*}

For $k>0$, we have again by recurrence on the size of the
first argument :
\begin{align*}
 F_{k+1}(0_{k+1},v,w) & = w \\
 F_{k+1}(S_{k+1}(t),v,w) & = F_k(v,v,F_{k+1}(t,v,w)) \\
         & = g^{n^k}(F_{k+1}(t,v,w)) && \text{by recurrence on $k$} \\
         & =  g^{n^k}( g^{|t| \times n^k}(w)) && \text{by recurrence
           on $t$} \\
         & = g^{(|t|+1)n^{k}}(w)\rlap{\hbox to216 pt{\hfil}\qEd}
\end{align*}
%end{proof}

The sequence of functions $(F_k)_{k}$ allows us to define polynomial
length iterators over $\sW(0)$.
\begin{lem}\label{uniPoly}
Let $P[X]$ be a polynomial of degree $k$ with natural coefficients and 
 $g:\sW(0)^p \sra \sW(0)^p$.
There is a function $\tilde{P} : \sW(k),\sW(0)^p \sra
\sW(0)^p$ in $\SRI{\sW}{k}$
such that for each $x$ and $y$, 
\begin{align} \label{eq:polIter}
\tilde{P}(x,y) & = g^{P(|x|)}(y)
\end{align}
\end{lem}

\proof%begin{proof}
The proof is done by induction on the degree of the polynomial.
The base case is trivial.
Suppose that the degree of $P$ is $k+1$. 
Hence, $P(x) = c.x^{k+1} + Q(x)$ where the degree of $Q$ is less or equal to $k$.
Suppose that $\tilde{Q}$ satisfies the induction hypothesis wrt $Q$.
We define $T_{k+1}^{c}$ by composition as follows
\begin{align*}
T_{k+1}^{0}(x,y) & = \tilde{Q}(x,y) \\
T_{k+1}^{d+1}(x,y) & = F_{k+1}(x,x,T_{k+1}^{d}(x,y))
   && d<c
\end{align*}
We set $\tilde{P}(x,y) = T_{k+1}^{c}(x,y)$.
Here $T_{k+1}^{c}(x,y)$ is defined by $c$ compositions of $F_{k+1}$ where $c$ is given and fixed.

We show by an induction on $c$ that $\tilde{P}(x,y)$
satisfies~\ref{eq:polIter}. We just show the inductive step below.
\begin{align*}
\tilde{P}(x,y) & = T_{k+1}^{d+1}(x,y) = F_k(x,x,T_{k+1}^{d}(x,y))  && \text{by dfn}\\
   & = F_k(x,x,g^{d.n^{k+1}+Q(n)}(y)) && \text{where $n=|x|$} \\
   & = g^{n.n^k}(g^{d.n^{k+1}+Q(n)}(y)) && \text{by Lemma~\ref{prop:Fk}}
    = g^{(d+1).n^{k+1}+Q(n)}(y)\rlap{\hbox to11 pt{\hfill}\qEd}
\end{align*} 
%end{proof}

\begin{lem}
Any polynomial $P[X]$ with natural coefficients is represented in strict ramified arithmetics.
\end{lem}

\proof%begin{proof}
We set $\overline{P}(x) = \tilde{P}(x,\mv{0})$ in which  we replace $g$ by the
successor $\suca{0}$. So, we have $\overline{P}(x)=\suca{0}^{P(x)}(\mv{0})$.
\qed%end{proof} 

We say that a  multivariate polynomial $P[X_1,\ldots,X_n]$ with $n$ distinct
variables is simple if each monomial of $P$ is of the form $c.X_i^d$ for some
natural constants $c$ and $d$. For example $2x^2+3.y^2+4y$ is
simple, but $2yx^2+y$ is not. The degree of a simple polynomial is the
greatest exponent of $P$'s variables.

\begin{lem}\label{multiPoly}
Let $P[X_1,\ldots,X_n]$ be a simple polynomial of degree $k$
and
 let $g:\sW(0)^p \sra \sW(0)^p$.
There is a function $\tilde{P} : \sW(k)^{n},\sW(0)^p \sra
\sW(0)^p$ in $\SRI{\sW}{k}$
such that
for each $x_1,\ldots,x_n$ and $y$, 
\begin{align} \label{eq:polIterN}
\tilde{P}(x_1,\ldots,x_n,y) & = g^{P(|x_1|,\ldots,|x_n|)}(y)
\end{align}
\end{lem}

\proof%begin{proof}
The proof is done by induction on the number $n$ of variables.
The base case is a consequence of Lemma~\ref{uniPoly}.
Suppose that the simple polynomial $P$ has $n+1$ variables
$X_1,\ldots, X_n, X_{n+1}$. 
Since $P$ is simple, we write it as the sum 
$P(X_1,\ldots,X_n,X_{n+1}) = P'(X_1,\ldots,X_n) + P''(X_{n+1})$.
Suppose that $\tilde{P'}$ ($\tilde{P''}$) satisfies the induction
hypothesis wrt $P'$ (resp. $P''$).
We define $\tilde{P}$ by
\begin{align*}
\tilde{P}(x_1,\ldots,x_{n+1},y) & = \tilde{P'}(x_1,\ldots,x_{n},\tilde{P''}(x_{n+1},y))
\end{align*}

Indeed, we have 
\begin{align*}
\tilde{P}(x_1,\ldots,x_{n+1},y) & =
\tilde{P'}(x_1,\ldots,x_{n},g^{P''(|x_{n+1}|)}(y)) 
 = g^{P'(|x_1|,\ldots,|x_n|)}(g^{P''(|x_{n+1}|)}(y)) \\
 & = g^{P'(|x_1|,\ldots,|x_n|) + P''(|x_{n+1}|)}(y) = g^{P(|x_1|,\ldots,|x_{n+1}|)}(y)
 \rlap{\hbox to76 pt{\hfill}\qEd}
\end{align*}
%end{proof}

\subsection{Characterizing a polynomial time hierarchy}

\begin{thm}\label{thm:dtime}
The set of functions $\SRI{\sW}{k}$ is exactly $\DT(n^k)$. \\
That is  $\SRI{\sW}{k}=\DT(n^k)$.
\end{thm}

The demonstration of Theorem~\ref{thm:dtime} is a consequence of
Lemma~\ref{lem1} and~\ref{lem2} below.

\begin{lem}\label{lem1}
Let $\phi : \sW^{p} \sra \sW^{q}$ be a function which is
computable by a register machine $M$ in time $(c.\sum_{i=1,p}
n_i^k)+d$  for some
constants $c$,$d$ and $k$,
where $n_i$ is the size of the $i$\textit{th} argument.
Then, there is a function 
$f:\sW(k)^{p} \sra \sW(0)^{q}$ of $\SRI{\sW}{k}$ 
 such that 
for each $u$, $f(u) = \phi(u)$.
\end{lem}

\proof%begin{proof}
A configuration of $M$ is encoded by a $m+1$-uplet of $\sW(0)$ which
represents the state and the value of the $m$ registers of $M$. 
Then, it is not difficult
to design a function 
$\textit{next}:\sW(0)^{m+1} \sra  \sW(0)^{m+1}$, 
which given a configuration, produces the next
configuration wrt $M$.
The function $\textit{next}$ is based on nested flat recursions over $\sW(0)$.
To illustrate the construction of $\textit{next}$, consider that the register machine $M$ has just two registers $R_1$ and $R_2$.
We define the function \textit{next} for each state of $M$ by using flat recursion
in order to match a state and to switch to the right transition. 
The next configuration depends on the finite function $\lab$ of $M$.
For example if in state $s_i$, the value of $R_2$ is replaced by the value of $R_1$, and the next state is $s_j$, 
we define $\textit{next}$ by flat recursion such that
$
\textit{next}(s_i,R_1,R_2) = (s_j,R_1,R_1) 
$.

Now, we have to iterate $\textit{next}$ within the polynomial time bound.
For this we use Lemma~\ref{multiPoly} since it is a simple polynomial.

Therefore, there is a function 
$\textit{loop}:\sW(k)^{p}, \sW(0)^{m+1} \sra\sW(0)^{m+1}$
such that
\[\textit{loop}(x_1,\ldots,x_{p},\vec{y}) 
   = \textit{next}^{(c.\sum_{i=1,p} |x_i|^{k}) + d}(\vec{y})\ .
\]
We conclude by taking 
$f(x_1,\ldots,x_{p}) = \theta(\textit{loop}(x_1,\ldots,x_{p},\textit{init}))$,
where $\theta$ is a composition of projections and $\textit{init}$ is
the initial configuration, that is $\textit{init} = ( \textsc{begin},x_1,\ldots,x_{p},\mv{0},\ldots,\mv{0})$.
\qed%end{proof}

\begin{lem}\label{lem2}
Assume that $f:\sW(k_1)^{p_1},\ldots,\sW(k_n)^{p_n} \sra \sW(r)^{q}$ 
is in $\SRI{\sW}{k}$.
Then there is a polynomial $P$ of degree $k$, or less, 
such that for any $u_1,\ldots,u_n$, the computation of
$f(u_1,\ldots,u_n)$,  on register machines, is performed in time bounded by
$$
P(\max\{ |u_i| \,|\, \text{where the tier of $u_i$ is greater than
  $0$, that is $k_i > 0$}\}_{i=1,n})
$$
\end{lem}

\proof%begin{proof}
The proof goes by induction on $k$.
Suppose that $f \in \SRI{\sW}{0}$. In this case, the definition of
$\SRI{\sW}{0}$ claims that $f$ is not defined by
strict ramified recursion.  Hence, it is not hard to
compute $f$ in constant time. 
%In particular, this case includes all
%the cases where the output tier $r$ is strictly greater than $0$.

Now, suppose that $f \in \SRI{\sW}{k+1}$.  
There are two main cases that we are considering below.

First, $f$ is obtained by $k+1$-ramified recursion. 
We compute a loop whose length is bounded by the length of the first
argument $u_1$. 
We begin by evaluating $v_0 = h_{\epsilon}(u_2,\ldots,u_n)$. 
Next we compute $h_{\alpha}(u_2,\ldots,u_n,v_0)$ where $\alpha$ is the last
letter of $u_1$. And, we repeat this process till we have consumed all
letters of the recursion argument $u_1$. 
As usual with tiering system, the key point is that the runtime of the
auxiliary functions $h_{a}$ 
and $h_{b}$ does not depend on tier $0$ values. 
Hence we associate three polynomials
$P_{\epsilon}$, $P_a$ and $P_b$ 
satisfying the induction hypothesis. 
The runtime of $f$ is bounded by 
$P_{\epsilon}(\max\{ |u_i| \,|\, \text{where $k_i > 0$}\})) 
    + |u_1| \times \max_{\alpha=a,b}(P_{\alpha}(\max\{ |u_i| \,|\,
    \text{where $k_i > 0$}\}))$.
Since $ h_{\epsilon}$,$h_a$, and $h_b$ have domains which have
strictly lower tiers than $k+1$, it follows that degrees of the
corresponding polynomials, $P_{\epsilon}$, $P_a$ and $P_b$ are at most
$k$ by induction hypothesis. As a consequence, there is a polynomial which bounds 
$P_{\epsilon}(X)+X.\max_{\alpha=a,b}(P_{\alpha}(X))$
of degree at most $k+1$. This polynomial is an upper bound on $f$'s runtime.

Second, $f$ is defined by composition.
Say that $f(\vec{x}) = h(\vec{x},g(\vec{x}))$. 
There are two cases to consider. The first is when the output tier of
$g$ is $0$. In this case, the runtime of $f$ is bounded by the sum of
the runtime of $g$ and $h$. 
The second is when the output tier of $g$ is strictly greater than
$0$. Then, the runtime of $g$ is constant because $g$ cannot be
defined by recursion. It follows that the runtime of $f$ is bounded
by the runtime of $h$ plus an additive constant (due to $g$).
\qed%end{proof}

\section{Diagonalization with dependent tiers}
In this section, we consider again the sequence $(F_k)_k$
parameterized by a strictly increasing function $g$. Recall that, $F_k$ iterates $n^k$
times a function $g$ and is in $\SRI{\sW}{k}$.
Each function of $\SRI{\sW}{k}$ is eventually dominated by composition
of $F_k$ at tier $0$. But, $F_k$ is not in $\SRI{\sW}{k+1}$. 
This leads us to ask two questions: How to jump from $\SRI{\sW}{k}$
to $\SRI{\sW}{k+1}$? And how to jump outside $\cup_k \SRI{\sW}{k}$ ?
In other words, this leads us to investigate jump operators, which allows to 
define $(F_k)_k$ sequence of functions by iteration and to diagonalize it in order to compute a function, which is not in $\cup_k \SRI{\sW}{k}$.

\subsection{Jumping from $\mathbf{\SRI{\sW}{k}}$ to $\SRI{\sW}{k+1}$}\label{sec:jump}
In order to answer to the first question, we introduce an operator
$\itd{\_}$ such that for each $k$,
$\itd{F_k}:\sW(k+1),\sW(k),\sW(0) \sra \sW(0)$ and
\begin{align*}
\itd{F_k}(\mv{k+1},x,y) & = y \\
\itd{F_k}(\suc{k+1}(r),x,y) & = F_k(x,x,\itd{F_k}(r,x,y))
\end{align*}
From definitions, it is clear that $\itd{F_k}(r,x,y) = F_{k+1}(r,x,y)$. 
Observe also, that the operator $\itd{\_}$ respects the strict
ramification principle.

\begin{defi}
Let $h:\sW(k_1)^{p_1},\ldots,\sW(k_n)^{p_n} \sra \sW(r)^{q}$ 
be an $n$-placed function and let $f:\sW(k) \sra \sW(0)$ be a
$1$-placed function.
We say that $h$ is dominated by $f$ if $|h(x_1,\ldots,x_n)| \leq |f(x)|$
holds for all $x_1,\ldots,x_n$ and $x$ with $x_1 \leq x, \ldots, x_n \leq x$.
\end{defi}

\begin{prop}
Each function $h$ of $\SRI{\sW}{k}$ is dominated by 
$f_k(x)=\itd{F_k}(a,x,b)$ for some $a\in\sW(k+1)$ and $b\in\sW(0)$. 
\end{prop}

\proof%begin{proof}
Since $\mathbf{\SRI{\sW}{k}} = \DT(n^k)$, 
there is $a'$ and $b'$ such that for all $x_1,\ldots,x_n$,
\begin{align*}
|h(x_1,\ldots,x_n)| & \leq a'(\sum |x_i|^k) + b'
\end{align*}
Let  $a=\suca{k+1}^{a'}(\mv{k+1})$ and $b=\suca{0}^{b'}(\mv{0})$
be two words such that $|a|=a'$ and $|b|=b'$.
It follows that for all $x_1,\ldots,x_n$ and $x$ with $x_1 \leq x, \ldots, x_n \leq x$, we have
$|h(x_1,\ldots,x_n)| \leq \itd{F_k}(a,x,b)$.
Indeed,
\begin{align*}
|h(x_1,\ldots,x_n)| & \leq |g^{|a|.|x|^k}(b)| && \text{since $g$ is assumed strictly monotonic} \\
                              & \leq |F_{k+1}(a,x,b)| && \text{by Lemma~\ref{prop:Fk}} \\
                              & = |\itd{F_k}(a,x,b)| && \text{by dfn}
\end{align*}
\qed%end{proof}

But, the important point here is that an operator like  $\itd{\_}$
allows to escape $\SRI{\sW}{k}$ because $\itd{F_k}=F_{k+1}$ is not in
$\SRI{\sW}{k}$.  
We now iterate $\itd{\_}$ starting from $F_0$ in order to
produce the chain of monotonic $F_k$ functions, as follows :
\begin{align*}
\itdk{0}{F_0}(r,x,y) & = F_0(r,x,y) \\
\itdk{k+1}{F_0}(r,x,y) & = \itd{\itdk{k}{F_0}}(r,x,y)
\end{align*} 
We say that the $k$\textit{th} iterate of $F_0$ is $\itdk{k}{F_0}$.

\begin{prop}\label{prop:itdkF0FK}
For all $k \in \N$, $r \in \tW(k)$, $x \in \tW(k)$ and $y \in \tW(0)$, we have
\begin{align*}
\itdk{k}{F_0}(r,x,y) & = F_k(r,x,y) 
\end{align*} 
\end{prop}

\proof%begin{proof}
The proof goes by induction on $k$. The base case is immediate.
Next,
\begin{align*}
\itdk{k+1}{F_0}(r,x,y) & = \itd{\itdk{k}{F_0}}(r,x,y) && \text{by dfn} \\
    & = \itd{F_{k}}(r,x,y) && \text{Ind. Hyp.} \\
    & = F_{k+1}(r,x,y) && \text{by dfn} \rlap{\hbox to78 pt{\hfill}\qEd}
\end{align*}
%end{proof}
Therefore, the $k+1$\textit{th} iterate of $F_0$ is in $\SRI{\sW}{k+1}$ but not in $\SRI{\sW}{k}$. 

\begin{rem}
The jump operator $\itd{\_}$ can be applied to any function of type
\[\sW(k),\sW(\max(k-1,0)),\sW(0) \sra \sW(0)\ .\] 
\end{rem}

\subsection{Jumping outside $\mathbf{\cup_k \SRI{\sW}{k}}$}\label{sec:bigjump}
\ \\
We define next a $4$-placed operator $\itdiag$ based on a double
recursion. It is a nested recursion based on lexicographic ordering.
\begin{align*}
\itdiag[g](0,r,x,y) & = g(y) && g:\sW(0)\sra \sW(0)\\
\itdiag[g](k+1,\mv{k+1},x,y) & = y \\ %&& \text{where $\suc{~}=\suca{~},\sucb{~}$}\\
\itdiag[g](k+1,\suc{k+1}(r),x,y) & = \itdiag[g](k,x,x,\itdiag[g](k+1,r,x,y))
\end{align*}
Here, $\itdiag$ is parameterized by $g$.

\begin{prop}\label{prop:itdk}
For all $k$,$r$,$x$ and $y$, we have
\begin{align*}
\itdiag[g](k,r,x,y) & = \itdk{k}{F_0}(r,x,y)  
\end{align*}
\end{prop}

\proof%begin{proof}
By induction on $k$ and $r$.
\qed%end{proof}

If we fix the first argument $k$, we iterate on the second argument
$r$ of tier $k$ and we compute $F_k$.
Now, if we fix the second argument $r$, we jump from tier to tier which
allows to get outside each function set $\SRI{\sW}{k}$, computing the
successive iterate of $F_0$.
So, $\itdiag$ allows us to jump outside each $\SRI{\sW}{k}$ for any
$k$.

\begin{prop}
The $4$ placed function $\itdiag[\suca{0}]$ is not in $\cup_{k\in \N} \SRI{\sW}{k}$.
\end{prop}

\proof%begin{proof}
We set $\phi(x) = \itdiag[\suca{0}](|x|,x,x,x)$ for all $x$.
We have
\begin{align*}
\phi(x) & = \itdiag[\suca{0}](|x|,x,x,x) \\
           & = \itdk{|x|}{F_0}(x,x,x) && \text{by Prop.~\ref{prop:itdk} where $g=\suca{0}$} \\
           & = F_{|x|}(x,x,x) && \text{by Prop.~\ref{prop:itdkF0FK}} \\
           & = \suca{0}^{|x|^{|x|}}(x) && \text{by Prop.~\ref{prop:Fk} when $|x|>0$}          
\end{align*}

We see that $|\phi(x)| =  |x|^{|x|+1} + |x|$ which is clearly not in  
$\cup_{k \in \N} \SRI{\sW}{k}$ in which each function is polynomially bounded
as it has been established in  Theorem~\ref{thm:dtime}.
\qed%end{proof}

The operator $\itdiag$ produces a function, which is not in $\cup_{k\in \N}
\SRI{\sW}{k}$. That is, $\itdiag[g]$ is not a ramified function in $\cup_{k\in \N}\SRI{\sW}{k}$ if $g$ is increasing. 
However, we may see that intuitively the ``domain'' depends on the
first argument, and so we should write 
$\itdiag[g]:\forall k \in \N. \sW(k),\sW(k),\sW(0) \sra \sW(0)$. 
To formalize this idea, we now introduce a typed lambda-calculus
with very restricted dependent types and arithmetical gadgets.

\section{An applied lambda-calculus with dependent types}
%The aim of this last section is to present a typed lambda-calculus in which
%each function of $\SRI{\sW}{k}$ for any $k$ is representable from
%an iterator $\diag$ which mimics $\itdiag$.

\subsection{Types, terms, and rules}
We propose a typed $\lambda$-calculus $\LD$ in
which types depend on tiers. 
For this, we have a base type $\atier$ to denote tiers and a unary
predicate $\tW$ of kind $\atier \kdi \bkd$, which is intended to name words
at each tier.  

Raw expressions, Kinds, types and terms, are defined following the grammar rules :
$$
\begin{array}{llp{3mm}l}
\textit{(Type constructors)} & \alpha & ::=&\quad \atier \ | \ \tW \\
\textit{(Kinds)} & \kindterm  & ::= &\quad 
\bkd \ | \ \typone \kdi \kindterm \\
\textit{(Types)} & \typone & ::= &\quad \alpha
   \ | \ \typone \times \typone
   \ | \ \forall x. \typone\ | \ \typone\ \termone
\\
\textit{(Term constructors)} & \conone & ::=&\quad \tzero \ | \ \tsuc
\ | \ \tmv \ |\ \tsuca \ |\ \tsucb \ |\ \tpair{\_}{\_}{\_} \ |\ \tpra \ |\ \tprb \ |\ \tflat\ |\ \diag \\
\textit{(Terms)} & \termone & ::= &\quad
\conone \ | \ \varone \ | \  (\termone \termone) \ | \ \lambda x. \termone
\end{array}
$$ 
where $\varone$ is a variable.

The types assigned to type and term constructors are given in
Figure~\ref{fig:constants}. We may omit some brackets of a type or of
a term using familiar Currying conventions.

\begin{figure}%[ht]
\hrulefill\\ 
\begin{flushleft}
\textit{Type constructors}
\end{flushleft}
\begin{gather*}
\imp \atier : \bkd \\   
\imp \tW : \atier \Rightarrow \bkd
\end{gather*}
\begin{flushleft}
\textit{Terms of type $\atier$}
\end{flushleft}
\begin{gather*}
\imp \tzero:\atier   \\ 
\imp \tsuc:\atier \sra \atier
\end{gather*}
\begin{flushleft}
\textit{Tiered words}
\end{flushleft}
\begin{gather*}
\imp \tmv:\forall k. \tW(k) \\
\imp \tsuca:\forall  k. \tW(k)\sra \tW(k) \\
\imp \tsucb:\forall  k. \tW(k)\sra \tW(k)
\end{gather*}
\begin{flushleft}
\textit{Pairing and projections}
\end{flushleft}
\begin{gather*}
\imp \tpair{\_}{\_}{\_}:\forall k. \tW(k) \sra \tW(k) \sra
\tW(k)\times\tW(k) \\
\imp \tpra:\forall k. \tW(k)\times\tW(k)\sra\tW(k) \\
\imp \tprb:\forall k. \tW(k)\times\tW(k)\sra\tW(k)
\end{gather*}
\begin{flushleft}
\textit{Flat recursion}
\end{flushleft}
\begin{gather*}
\imp \tflat_{\tau}:\forall k. \tau \sra (\tW(k)\sra\tau)^2 \sra \tW(k) \sra\tau
\end{gather*}
\begin{flushleft}
\textit{Double tiered recursion}
\end{flushleft}
\begin{gather*}
\imp \diag_{p}:(\tW(\tzero)^p\sra\tW(\tzero)^p) \sra \forall k.  
 \tW(k) \sra \tW(k) \sra \tW(\tzero)^p \sra \tW(\tzero)^p
\end{gather*}
\caption{Types of type and term constructors \label{fig:constants}}
\hrulefill
\end{figure}

A term $\termone$ is of type $\tau$, that we write $\termone:\tau$, if
there is a derivation of $\imp \termone:\tau$ following the typing
rules of Figure~\ref{fig:typingRules}. We note $\dom(\typenv)$ the set
of (term) variables declared in $\typenv$.
\begin{figure}[ht]
\hrulefill\\ 
\begin{flushleft}\textit{Kinding rules}\end{flushleft}
\begin{gather*}
\ninfer{\typenv \imp \phi:\tau \Rightarrow \kappa 
 \qquad \typenv \imp t:\tau} 
{\typenv \imp \phi t: \kappa}
{\text{$\Rightarrow$ elim}} \\[.2 cm]
\end{gather*}
%%%%
\begin{flushleft}
\textit{Typing rules}
\end{flushleft}
\begin{gather*}
\ninfer{\typenv \imp \tau:\bkd}{\typenv,x:\tau \imp x:\tau} 
{\text{Variable, $x \not \in \dom(\typenv)$}} \\[.2 cm]
%%%%
\ninfer{ }{\typenv \imp \textbf{c}:\tau} 
{\text{where $\textbf{c}$ is a type or a term constructor of type $\tau$}} \\[.2 cm]
%%%%
\ninfer{\typenv, x:\tau \imp \termone:\sigma}
{\typenv \imp \lambda x. \termone:\tau \sra \sigma} 
{\text{$\sra$ intro}} \\[.2 cm]
%%%%%
\ninfer{\typenv \imp \termone:\tau \sra \sigma \qquad \typenv \imp \termtwo:\tau} 
{\typenv \imp \termone\termtwo:\sigma}
{\text{$\sra$ elim}} \\[.2 cm]
%%%%%
\ninfer{\typenv,x:\atier \imp \termone:\tau}
{\typenv \imp \lambda x. \termone:\forall x. \tau}
{\text{$\forall$ intro, and $x \not \in \dom(\typenv)$}} \\[.2 cm]
%%%%%
\ninfer{\typenv \imp \termone:\forall x. \tau  \qquad \typenv \imp k:\atier} 
{\typenv \imp \termone k:\tau[x \leftarrow k]}
{\text{$\forall$ elim}} \\[.2 cm]
%\ninfer{\imp \tau:\bkd \qquad \typenv \imp \phi:\kappa}{\typenv,x:\tau \imp \phi:\kappa} 
%{\text{Weakening, $x\not \in \dom(\typenv)$}}  \\[.2 cm]
\end{gather*}
%%%%
\begin{flushleft}
\textit{Weakening rule}
\end{flushleft}
\begin{gather*}
\ninfer{\imp \tau:\bkd \qquad \typenv \imp M:\sigma}{\typenv,x:\tau \imp M:\sigma} 
{\text{Weakening, $x\not \in \dom(\typenv)$ and $M$ is a term or a type}} \\[.2 cm]
%%%%%%
%\ninfer{\typenv \imp \termone:\tau \qquad \typenv \imp \termtwo:\sigma}
%{\typenv \imp \langle \termone,\termtwo \rangle : \tau \times \sigma}
%{\text{$\times$ intro}} \\[.2 cm]
%%%%%%
%\ninfer{\typenv \imp \termone: \tau \times \sigma}
%{\typenv \imp \tpra \termone:\tau}
%{\text{$\times$ elim left}} \\[.2 cm]
%%%%%%
%\ninfer{\typenv \imp \termone: \tau \times \sigma}
%{\typenv \imp \tpra \termtwo:\sigma}
%{\text{$\times$ elim right}} \\[.2 cm]
%%%%% 
\end{gather*}
%%%%
\begin{flushleft}
\textit{Downcasting rule}
\end{flushleft}
\begin{gather*}
\ninfer{\typenv \imp \termone:\tW(\tsuc(t))} 
{\typenv \imp \termone:\tW(t)}
{\text{Downcasting}}
\end{gather*}
\caption{Typing rules.\label{fig:typingRules}}
\hrulefill
\end{figure}

The one step (contextual) reduction $\red$ is defined in Figure~\ref{fig:computation}.
The transitive closure of $\red$ is $\redtr$. Here $M[x\leftarrow N]$
means the usual substitution of all free occurrences of $x$ in $M$ by $N$.

\begin{figure}[ht]
\hrulefill\\ 
\begin{flushleft}
$\beta$-reduction
\end{flushleft}
\begin{gather*}
(\lambda x. M) N \red M[x\leftarrow N] 
\end{gather*} 
\begin{flushleft}
projections
\end{flushleft}
\begin{gather*}
(\tpra k\ \tpair{M}{N}{k})  \red M \\
(\tprb k\ \tpair{M}{N}{k})  \red N 
\end{gather*} 
\begin{flushleft}
flat recursion
\end{flushleft}
\begin{gather*}
(\tflat\ k\ h_\epsilon \ h_a\ h_b\ (\tmv\ k)) \red h_{\epsilon} \\
(\tflat\ k\ h_\epsilon \ h_a\ h_b\ (\tsuca\ k\ x)) \red (h_{a}\ x)\\
(\tflat\ k\ h_\epsilon\ h_a \ h_b\ (\tsucb\ k\ x)) \red (h_{b}\  x)
\end{gather*} 
\begin{flushleft}
double recursion where $\tsucj=\tsuca,\tsucb$
\end{flushleft}
\begin{gather*}
(\diag_p\ g\ \tzero\ z \ x \ y)  \red (g\ y)  \\
(\diag_p\ g\ (\tsuc\ k)\ \tmv\ x\ y)  \red y \\% && \text{where}\\
(\diag_p\ g\ (\tsuc\ k)\ (\tsucj \ k\ r)\ x\ y)  \red (\diag_p\ g\ k\ x\ x\
(\diag_p\ g\ (\tsuc\ k)\ r\ x\ y)) 
\end{gather*} 
\caption{Rules of computation \label{fig:computation}}
\hrulefill
\end{figure}

\begin{rem}\hfill
\begin{enumerate}[(1)]
\item As usual, $\tau \Rightarrow \kappa$ and $\tau \sra \tau'$ are
  short cuts for $\Pi x:\tau.\kappa$ and $\forall x:\tau.\tau'$,
  when $x$ is not occurring in $\kappa$ or $\tau'$.
\item There is no type variable (except type constructors).
\item In fact, we just consider two kinds $\bkd$ and $\atier
  \Rightarrow \bkd$, because we have no introduction rules for kinds.
\item The two previous points imply that in a judgment of the form
  $\typenv \imp \termone:\sigma$, if $(x:\tau)$ is in $\termone$,
 then $\tau$ is either $\atier$ or $\tW(t)$ for some term $t$ of type $\atier$.
\end{enumerate}
\end{rem}

The system $\LD$ can be translated in the system T of G\"{o}del and so it
has the Church-Rosser
and strong normalization properties.

\subsection{Function representation at a given tier}\ \\

A natural number $k$ is represented by $\tr{k}$ thus:
\begin{align*}
\tr{0} & = \tzero & \tr{x+1} & = (\tsuc\ \tr{x})
\end{align*}
And a word $u$ of $\sW$ is represented by $\trw{k}{u}$ at tier $k$ thus
\begin{align*}
\trw{k}{\epsilon} & = (\tmv\  \tr{k}) & 
\trw{k}{a(x)} & = (\tsuca\ \tr{k} \ \trw{k}{x}) &
\trw{k}{b(x)} & = (\tsucb\ \tr{k} \ \trw{k}{x})
\end{align*}

%\begin{defi}
%Let $\phi:\sW^{p}\sra\sW^{q}$. The function $\phi$ is
%represented at tier $k$ if there is a term 
%$M:\tW(\tr{k_1})^{p_1} \sra \ldots \sra \tW(\tr{k_n})^{p_n} \sra \tW(\tr{0})$
%such that $p=p_1+\ldots+p_n$ and for all $u_1 \ldots u_{p} $ of $\sW^{p}$.
%\begin{align*}
%M \trw{k_1}{u_1} \ldots \trw{k_1}{u_{p_1}} \ldots
%\trw{k_n}{u_{\sum_{i<n} p_i +1 }}
%\ldots \trw{k_n}{u_{p}}    
%        & \redtr \trw{0}{\phi(u_1 \ldots u_{p})} 
%\end{align*}
%where $k=k_1$ and $k_1 > \ldots > k_n$.
%We define $\CI{k}{\sW}$ as the set of functions which are represented
%at tier $k$.
%\end{defi}

\begin{defi}
Let $\phi:\sW^{p+p'}\sra\sW^{q}$. The function $\phi$ is
represented at tier $k$ if there is a term 
$M:\tW(\tr{k})^{p} \sra \tW(\tr{0})^{p'} \sra \tW(\tr{0})^{q}$
such that for all $u_1 \ldots u_p$ of $\sW^{p}$
and for all $v_1 \ldots v_{p'}$ of $\sW^{p'}$
\begin{align*}
M \ \trw{k}{u_1} \ldots \trw{k}{u_{p}} \ \trw{0}{v_1} \ldots \trw{0}{v_{p'}}   
        & \redtr \trw{0}{\phi(u_1,\ldots,u_{p},v_1,\ldots,v_{p'} )} 
\end{align*}
We define $\CI{k}{\sW}$ as the set of functions which are represented
at tier $k$.
\end{defi}

\begin{lem}\label{lem:CL0}
Each function $g:\sW(0)^p \sra \sW(0)^q$ in $\SRI{\sW}{0}$ is represented at
tier $0$, and so is in $\CI{0}{\sW}$.
\end{lem}

\proof
The proof is done by induction on the definition of $g$.
\qed

The construction of a polynomial length iterator in $\LD$ follows closely the lines
of the demonstration of Lemma~\ref{multiPoly}. It is obtained by composition from $F_k$ functions, which are
representable at tier $k$ following the Lemma below.

\begin{lem}
For each $k$, the function $F_k$ parameterized by a function $g:\sW(0)^p \sra \sW(0)^p$,
is represented at tier $k$, and so  is in $\CI{k}{\sW}$.
\end{lem}

\proof%begin{proof}
The previous lemma~\ref{lem:CL0} gives 
a term $N:\tW(\tzero)^p \sra \tW(\tzero)^p$,
which represents $g$.
Now, for each $k$, we define a sequence of terms $(M_{k})_k$ parameterized by $N$ by
\begin{align*}
M_{k} & = \lambda x \lambda y \lambda z. (\diag_p\ N\ \tr{k}\ x\
y\ z)
&&
\text{of type $\tW(\tr{k}) \sra \tW(\tr{k}) \sra \tW(\tzero)^p  \sra
  \tW(\tzero)^p$}
\end{align*}
We can check that 
$F_{k}$ is represented at tier $k$ by $M_{k}$ by induction on $k$ and the first parameter of $M_{k}$: \\
For $k=0$ and for all $u$,$v$ and $w$, we have
\begin{align*}
(M_{0} \ \trw{0}{u}\  \trw{0}{v}\ \trw{0}{w} ) & = (\diag_p\ N\ \tr{0}\ \trw{0}{u}\  \trw{0}{v}\ \trw{0}{w}) \\
          & \red (N\ \trw{0}{w}) \\
          & = \trw{0}{g(w)} = \trw{0}{F_{0}(u,v,w)}
\end{align*}
For $k+1$, we proceed by induction on the first argument $u$ of $M_{k+1}$.
First, for all $v$ and $w$, we have
\begin{align*}
M_{k+1} \ \trw{k+1}{\epsilon}\  \trw{k+1}{v}\ \trw{0}{w}  & = \diag_p\ N\ \tr{k+1}\ \trw{k+1}{\epsilon}\  \trw{k+1}{v}\ \trw{0}{w} \\
          & \red \trw{0}{w} = \trw{0}{F_{k+1}(\epsilon,v,w)} 
\end{align*}

\begin{align*}
M_{k+1} \ \trw{k+1}{i(u)}\  \trw{k+1}{v}\ \trw{0}{w}  & = \diag_p\ N\ \tr{k+1}\ \trw{k+1}{i(u)}\  \trw{k+1}{v}\ \trw{0}{w} && \text{$i=a,b$}\\
          & \red (\diag_p\ N \ \tr{k} \ \trw{k}{v}\ \trw{k}{v}\ (\diag_p\ N\ \tr{k+1}\ \trw{k+1}{u}\ \trw{k+1}{v}\ \trw{0}{w})) \\
          & = \trw{0}{F_{k}(v,v,F_{k+1}(u,v,w))} = \trw{0}{F_{k+1}(i(u),v,w)} 
\end{align*}
In conclusion, for all $k$, $u$, $v$ and $w$, we have
\begin{align*}
 M_k\ \trw{k}{u}\ \trw{k}{v}\ \trw{0}{w} & = \trw{0}{F_{k}(u,v,w)}
\end{align*}
It is also worth to see that by Lemma~\ref{prop:Fk}:
$$%\begin{align*}
 M_k\ \trw{k}{u}\ \trw{k}{v}\ \trw{0}{w}  = \trw{0}{g^{|u|.|v|^k}(w)}\eqno{\qEd}
$$% \end{align*}
%end{proof}

The following Lemma corresponds to Lemma~\ref{prop:Fk}
\begin{lem}\label{TypeduniPoly}
Any polynomial $P[X]$ of degree $k$ with natural coefficients is represented at tier $k$ in $\CI{k}{\sW}$.
More precisely, assume that $N:\tW(0)^p \sra \tW(0)^p$. \\
Then, there is a  term $P_k : \tW(k),\tW(0)^p \sra \tW(0)^p$ in $\CI{k}{\sW}$
such that for each $u$ and $v$, 
\begin{align}
P_k \ \trw{k}{u} \ \trw{0}{v} & \redtr (N^{P(|\trw{k}{u}|)}\ \trw{0}{v})
\end{align}
\end{lem}

\proof%begin{proof}
The proof goes by induction.
Suppose that the degree of $P$ is $k+1$. 
Hence, $P(x) = c.x^{k+1} + Q(x)$ where the degree of $Q$ is less or equal to $k$.
Suppose that $M'$ satisfies the induction hypothesis wrt $Q$.
We define $M_{k+1}^{c}$ by composition as follows
\begin{align*}
M_{k+1}^{0}\ x\ y & = (M' x\ y) \\
M_{k+1}^{d+1}\ x\ y & = (M_{k+1} \ x\ x \ (M_{k+1}^{d}\ x\ y))
   && d<c
\end{align*}
where $(M_{k})_k$ is the sequence of terms defined in the demonstration of the previous Lemma, and computes $(F_k)_k$.
The type of $M^{d}_{k+1}$ is $\tW(k+1) \sra \tW(0)^p \sra \tW(0)^p$ for any $d$.
We set $P_{k+1} = M_{k+1}^c$.
\qed%end{proof}

As a direct consequence of the above Lemma, we have a result
which is analogous to Lemma~\ref{multiPoly}:

\begin{cor}\label{cor:poly}
Let $P[X_1,\ldots,X_n]$ be a simple polynomial of degree $k$. \\
There is a term $\mathbf{P} : \tW(k)^{n},\tW(0)^p \sra \tW(0)^p$ such that
for each $u_1,\ldots,u_n$, and $v$, 
\begin{align*} \label{eq:polIterN}
(\mathbf{P} \ \trw{k}{u_1} \ldots \ \trw{k}{u_n}\ \trw{0}{v}) & \redtr (N^{P(|u_1|,\ldots,|u_n|)}\ \trw{0}{v})
\end{align*}
where $N:\tW(0)^p \sra \tW(0)^p$. 
\end{cor}

\proof%begin{proof}
The proof is done by induction on the number of variables.
The base case is a consequence of Lemma~\ref{TypeduniPoly}.
Suppose that the simple polynomial $P$ has $n+1$ variables
$X_1,\ldots, X_n, X_{n+1}$. 
Since $P$ is simple, we write it as the sum 
$P(X_1,\ldots,X_n,X_{n+1}) = P'(X_1,\ldots,X_n) + P''(X_{n+1})$.
Suppose that $\mathbf{P'}$ ($\mathbf{P''}$) satisfies the induction
hypothesis wrt $P'$ (resp. $P''$).
We define $\mathbf{P}$ by
$$%\begin{align*}
(\mathbf{P}\ x_1 \ldots\ x_{n+1})  = (\mathbf{P'}\ x_1 \ldots\ x_{n} (\mathbf{P''}\ x_{n+1}\ y))\eqno{\qEd}
$$%\end{align*}
%end{proof}

\begin{thm}
The set of functions $\SRI{\sW}{k}$ is exactly the set $\CI{k}{\sW}$,
that is the class $\DT(n^k)$.
\end{thm}

\proof%begin{proof}
First, we establish that $\DT(n^k) \subseteq \CI{k}{\sW}$. For this,
observe that the transition function $\textit{next}$, which is defined in the
proof of Lemma~\ref{lem1}, is represented at tier $0$, by a term of
type $\tW(0)^{m+1} \sra \tW(0)^{m+1}$. We iterate $\textit{next}$ by
using Corollary~\ref{cor:poly}.

Conversely, we show that $\CI{k}{\sW} \subseteq \SRI{\sW}{k}$. 
For this, let $f$ be a function represented at tier $k$ by a term $M$.
In other words, there is a normal derivation $\bigtriangledown$  
such that $\imp \termone:\tW(\tr{k})^p \sra \tW(\tr{0})^q$.
Observe that if $x$ is a variable of $M$, the type of $x$ is
$\tW(k')^{p'}$, $k' \leq k$ and $p'\leq p$.
So, a subterm $t$ of $M$ of type $\atier$ does not contain a variable
(of type $\atier$) 
and so represents a natural number, that is $t=\tr{r}$ for some $r$. 
Therefore, the term $M$ denotes a function of $\SRI{\sW}{k}$.
The proof is complete by Theorem~\ref{thm:dtime}. 
\qed%end{proof}

\subsection{Jumping outside}
Let $\phi:\sW \sra\sW$. The function $\phi$ is
represented at tier $\omega$ if there is a term 
$M:\forall k. \tW(k)^p \sra \tW(\tzero)^q$
such that
for all $u$,
\begin{align*}
M\ \tr{k}\ \trw{k}{u} \redtr \trw{0}{\phi(u)}
&& \text{where $k=|u|$}
\end{align*}
We define $\CI{\omega}{\sW}$ as the set of functions which are represented
at tier $\omega$.

\begin{proposition}
There is a function represented at tier $\omega$ which is not
representable at tier $k$, for any $k$. In other words, this function is not $\cup_k \SRI{\sW}{k}$.
\end{proposition}

\proof%begin{proof}
The function $\itdiag[\suca{0}]$ is representable at tier $\omega$. 
As the consequence, we can define the exponential as follows:
$E = \lambda k \lambda x.(\diag_1 (\tsuca\ \tzero)\ k\ x\ x\ x)$ of
type $\forall k. \sW(k) \sra \sW(0)$.
We have
  $E\ \tr{|u|}\ \trw{k}{u} \redtr e$
and $|e| \geq |u|^{|u|+1} + |u|$, for all $u$.
\qed%end{proof}

%\begin{figure}%[ht]
%\hrulefill\\ 
%\begin{flushleft}
%\textit{Type constructors}
%\end{flushleft}
%\begin{flushleft}
%\textit{Terms of type $\atier$}
%\end{flushleft}
%\begin{align*}
%\sem{\tzero} & = 0 & \sem{\tsuc}(x) & =\suc(x)
%\end{align*}
%\begin{flushleft}
%\textit{Tiered words}
%\end{flushleft}
%\begin{align*}
%\sem{\tmv \tr{k}} & = \mv{k} &
%\sem{\tsuca \tr{k}} x & = \suca{k}(x)
%&
%\sem{\tsucb \tr{k}} x & = \sucb{k}(x)
%\end{align*}
%\begin{flushleft}
%\textit{Pairing and projections}
%\end{flushleft}
%\begin{align*}
%\sem{\tpair{\tr{k}}{x}{y}} &=  &
%\sem{\tpra \tr{k} x} & = 
%\sem{\tprb \tr{k} x} & = 
%\end{align*}
%\begin{flushleft}
%\textit{Flat recursion}
%\end{flushleft}
%\begin{align*}
%\tflat_{\tau} k :\forall k. \tau \sra (\tW(k)\sra\tau)^2 \sra \tW(k)\sra\tau
%\end{align*}
%\begin{flushleft}
%\textit{Double tiered iteration}
%\end{flushleft}
%\begin{align*}
%\imp \diag_{alpha}:(\tW(\tzero)^p\sra\tW(\tzero)^p) \sra \forall k.  
% \tW(k),\tW(k),\tW(\tzero)^p \sra \tW(\tzero)^p
%\end{align*}
%\hrulefill
%\end{figure}
\subsection{Other ways to jump}
We have presented a manner of constructing an exponential function by
diagonalizing functions defined by strict ramified recursion. 
There are other approaches. In~\cite{Lei99}, Leivant ramifies the system T
of G\"odel~\cite{God58} by introducing an atomic type constructor
$\Omega(\tau)$ which allows to perform recursion over type $\tau$
terms. Thus, he obtains a characterization of $\FP$ and of the
elementary functions.

Bellantoni and Niggl~\cite{BN99} characterized the Grzegorczyk
hierarchy starting from the class $\FP$. For this, they define a rank
function which, roughly speaking, is a bound on the number of nested
recursions. The work of Caporaso, Covino and Pani seems also related
to the research presented in this paper, see~\cite{CCP01}.  We are
also aware of other related works like the one of Oitavem~\cite{Oit97}
or the one of Beckmann and Weiermann~\cite{BW96b}.  Finally,
Danner~\cite{Danner} proposed a ramified G\"odel system T with a
dependant typing system to study primitive recursive functions.

\vskip-20 pt

\end{document}